\documentclass{statsoc} 
\usepackage[a4paper]{geometry}
\usepackage[T1]{fontenc}
\usepackage[utf8]{inputenc}
\usepackage{graphicx}
\usepackage{amsmath}
\usepackage{natbib}
\usepackage{amssymb}
\usepackage{mathrsfs}

\usepackage[normalem]{ulem}

\usepackage{etoolbox}

\usepackage{color}

\makeatletter
\patchcmd{\@makecaption}
  {\parbox}
  {\advance\@tempdima-\fontdimen2} 
  {}{}
\makeatother

\definecolor{myred}{rgb}{1.0, 0.0, 0.0}

\newcommand{\red}[1]{\textcolor{myred}{#1}}

\newtheorem{theorem}{Theorem}

\title[The Latent Order Logistic Model]{A New Generative Statistical Model for Graphs: The Latent Order Logistic (LOLOG) Model}

\author[Ian E. Fellows]{Ian E. Fellows}
\address{Fellows Statistics, San Diego, United States}
\email{ian@fellstat.com}

\begin{document}

\maketitle

\begin{abstract}
Full probability models are critical for the statistical modeling of complex networks, and yet there are few general, flexible and widely applicable generative methods. We propose a new family of probability models motivated by the idea of network growth, which we call the Latent Order Logistic (LOLOG) model. LOLOG is a fully general framework capable of describing any probability distribution over graph configurations, though not all distributions are easily expressible or estimable as a LOLOG. We develop inferential procedures based on Monte Carlo Method of Moments, Generalized Method of Moments and variational inference. To show the flexibility of the model framework, we show how so-called scale-free networks can be modeled as LOLOGs via preferential attachment. The advantages of LOLOG in terms of avoidance of degeneracy, ease of sampling, and model flexibility are illustrated. Connections with the popular Exponential-family Random Graph model (ERGM) are also explored, and we find that they are identical in the case of dyadic independence. Finally, we apply the model to a social network of collaboration within a corporate law firm, a friendship network among adolescent students, and the friendship relations in an online social network.  
\end{abstract}

\section{Introduction}

As the information age continues to accelerate, the number and size of network datasets likewise grows. Online social networks, citation databases, and computer networks are larger and more accessible than ever. Yet there are few general, flexible and widely applicable generative statistical models that can be applied to these data.

Perhaps the most popular generative model in the social sciences is the (curved) Exponential-family Random Graph Model (ERGM) \citep{frank1986markov, hunter2006inference}. ERGMs have seen wide use across the social sciences \citep{van2016introduction}, and have inspired mathematical extensions modeling temporal networks \citep{guo2007recovering, krivitsky2014separable}, valued networks \citep{krivitsky2012exponential}, and networks with random nodal covariates \citep{fellows2012exponential}. While ERGMs are flexible and general, they have received criticism for their lack of scalability and propensity for model degeneracy \citep{chatterjee2013estimating,goldenberg2010survey}.

Additionally, there has been considerable attention paid in the literature to specific network generating procedures that can describe a particular feature observed in empirical networks. For example, \cite{barabasi1999emergence} explored the emergence and consequences of scale-free degree structures using a preferential attachment procedure. Other authors have explored features such as transitivity \citep{holme2002growing, bansal2009exploring}, emergent communities \citep{bianconi2014triadic}, densification \citep{leskovec2005graphs}, and alternate degree distributions \citep{krapivsky2000connectivity} to name a few. These generation procedures are often proposed without an inferential framework from which to estimate parameters from observed data, and sometimes don't have a fully specified probability model associated with them.

The contribution of this work is to propose a new general probabilistic framework for network analysis called the Latent Order Logistic (LOLOG) model. Section \ref{sec:lolog} develops the model specification and Sections \ref{sec:inf}, \ref{sec:inf2}, and \ref{sec:var} explore mechanisms for model estimation and inference. While ERGMs can be motivated with an appeal to the equilibrium distribution of a tie formation/dissolution process \citep{snijders2006new}, LOLOG is motivated by the principle of network growth. Despite this there is a strong connection between the two model classes, which is described in Section \ref{sec:ergm}.

As a consequence of its motivating data generation process, growth models such as the Barabasi-Albert procedure can be represented as a LOLOG with very few terms (see Section \ref{sec:scale}). Terms can be mixed and matched to provide a fuller representation of the more complex structures present in real graphs. In Section \ref{sec:ham} we perform one such analysis that simultaneously captures the degree, nodal covariate influences, transitivity and densification features of an online social network.

\section{Motivation from Considering the Hypothetical Growth of a Network}

Networks tend to evolve over time. This evolution can take the form of the addition or deletion of a tie from the network, or the addition or deletion of a vertex. To motivate our model we consider a growth process, where each edge variable is sequentially considered for edge creation, and edges are not deleted.

Let $Y^t \in \mathscr{Y}$ be a random $n$ by $n$ matrix representing a graph at time $t$ such that the edge variable $Y^t_{ij}$ is 1 if there is an edge from vertex $i$ to $j$, and 0 otherwise, where $\mathscr{Y}$ is the set of all possible graph configurations. Further let $y^t$ be a realization of the random variable $Y^t$. If the graph is undirected, then $Y^t$ is symmetric ($Y^t_{ij} = Y^t_{ji}$), and thus the dimensionality of $Y^t$ is $n_d=n(n-1)$ for directed graphs and $n_d={n(n-1)}/{2}$ for undirected graphs. 

Let $S \in \mathscr{S}$ be a random vector with realization $s$ of length $n_d$ indicating the order in which edge variables are considered for tie formation, where element $S_t=(i,j)$ indicates that the edge variable from vertex $i$ to $j$ is being considered at time $t$. $\mathscr{S}$ is the set of all possible orderings of edge variables, and so $S$ contains no repeated elements. The subscript $S_{\leq t}$ is used to denote the first $t$ elements of $S$. At initial formation, there are no connections between the $n$ vertices in the graph, so $Y_{ij}^0=0 \ \forall \ i,j \in \{1,...,n\} $. Edge variables remain set to 0 until they appear in $S$, so $Y_{ij}^t=0 \ \forall \ (i,j) \notin S_{\leq t}$. After they appear in $S$ edge variables may either be 0 or 1. We will sometimes denote the fully evolved graph as $Y=Y^{n_d}$.

Note that we have all edge variables existing in the graph at all time points, but they are set to 0 prior to being considered. We could have equivalently defined the edge variable set as growing such that edge variables don't exist prior to their appearance in $S$, and may be either 1 or 0 after. We find our choice of notation simpler.

A graph statistic is any function of the graph and order, $g(y^t,s_{\leq t})$, that is of interest to the modeler.  Without loss of generality, we will assume that the graph statistics are zero at network initialization ($g(y^0,s_0)=0$). These often take the form of subgraph counts (e.g., the number of edges, triangles or two-stars in the graph \citep[See][]{frank1986markov}). We define the change in the graph statistic when considering the $t$th edge variable ($y_{s_t}$) in the graph as
$$
c(y^t_{s_t} | y^{t-1}, s_{ \leq t}) = g(y^{t},s_{\leq t}) - g(y^{t-1},s_{\leq (t-1)}),
$$
and thus the graph statistic can be represented as the sum of the change statistics
$$
g(y,s) = \sum_{t=1}^{n_d} c(y_{s_t} | y^{t-1}, s_{ \leq t}).
$$
A statistic is order independent if the value of the statistic depends only on the graph, and not the order in which it was formed (that is, $g(y,s) = g(y)$). A statistic is called dyad independent if its change statistics do not depend on the rest of the graph (that is, $c(y_{s_t} | y^{t-1}, s_{ \leq t}) = c(y_{s_t}),\ \ \forall \ \  y_{s_t}, y^{t-1}, s$). A dyad independent statistic may depend on vertex covariates. For example, if a vertex represents a person, then the change statistic of an edge variable might depend on the age, sex or ethnicity of the two vertices it connects.

\section{The Latent Order Logistic (LOLOG) Formulation} \label{sec:lolog}



Our goal is to motivate a full probability framework for $S$ and $Y$ using network growth as the underlying data generating process. The growth process is determined by the distribution of the ordering $p(s)$. In our development we allow for any arbitrary distribution for $p(s)$; however, the choice of $p(s)$ should be dependent on the domain from which the data came. For example, $p(s)$ may be a complete random shuffle of edge variables. 

For other networks it may be reasonable to posit that vertices ``entered'' the network in some order that may or may not be random, and upon ``entering'' the network, edge variables connecting them to vertices already in the network are considered in a completely random order. Network growth models where vertices are added one-by-one have received considerable attention in the literature \citep{barabasi1999emergence, albert2002statistical, amaral2000classes}. The examples in Sections \ref{sec:l}, \ref{sec:add}, and \ref{sec:ham} all use this sequential vertex ordering process.

Suppose that the first $t-1$ edge variables have been considered previously, and we wish to model the probability of observing the next edge variable in the sequence ($y_{s_t}$) based on some set of graph statistics. We formulate the probability as a logistic regression using the change in the graph statistics as predictors
\begin{equation} \label{eq:condlolog}
p(y_{s_t} | \theta, y^{t-1}, s_{ \leq t}) = \frac{1}{Z_t}e^{\theta \cdot c(y_{s_t} | y^{t-1}, s_{ \leq t})},
\end{equation}
where the $\theta$ are the natural parameters and 
$$
Z_t = e^{\theta \cdot c(1 | y^{t-1}, s_{ \leq t})} + e^{\theta \cdot c(0 | y^{t-1}, s_{ \leq t})}
$$ 
is the normalizing constant (the dependence on $\theta, y^{t-1}$ and $s_{ \leq t}$ is suppressed). Logistic regression is a natural choice because it is an exponential family model and therefore maximum entropy \citep{barndorff1978information}.

The distribution of the graph conditional upon an observed sequence order is 
\begin{align}
p(y|s,\theta) &= \prod_{t=1}^{n_d} \frac{1}{Z_t}e^{\theta \cdot c(y_{s_t} | y^{t-1}, s_{ \leq t}) } \\
&=\frac{1}{\prod_{t=1}^{n_d}Z_t}e^{\theta \cdot g(y,s)} \ \ \ \ \ y \in \mathscr{Y} \ \ \textrm{and} \ \ s \in \mathscr{S}.
\end{align}
This distribution has several desirable properties. First, because $p(y|s,\theta)$ is the product of easily computable likelihoods, it is itself very easy to compute. Second, because it has exponential-family form, it has the maximum entropy property. Finally, it is very easy to sample from the joint distribution $p(y|s,\theta)p(s)$ as follows:
\begin{center}
  \uline{Drawing Independent Samples from a LOLOG Model}\\
\end{center}
\begin{enumerate}
    \item Set $y_{ij} = 0 \ \ \forall \ \ i, j \ \in \ \{1,...,n\}$.
    \item Draw a sample $s \sim p(S)$.
    \item For $t \ \in \ \{1,...,n_s\}$ draw and set $y_{s_t} \sim p(Y_{s_t} | \theta, y^{t-1}=y, s_{ \leq t})$.
\end{enumerate}
Provided each $c(\cdot| y^{t-1}, s_{ \leq t})$ can be calculated in constant time, generating an independent sample has $O(n^2)$ time complexity.


In practice the order in which edge variables are considered is not typically (fully) observed. The marginal exponential-family Latent Order Logistic model may be expressed as
\begin{equation} \label{eq:lolog}
p(y | \theta) = \sum_s p(y | s,\theta) p(s)  \ \ \ \ \ y \in \mathscr{Y} \ \ \textrm{and} \ \ s \in \mathscr{S}.
\end{equation}
The sum in Equation (\ref{eq:lolog}) is of high dimension, and thus the likelihood is intractable in general. However, if all terms are dyad independent, then $Y$ is independent of $S$, and thus $p(y | \theta)$ reduces to the product of logistic regression likelihoods. 



Two likelihood derivatives that will be useful in our subsequent developments are
{
\begin{align*}
    \frac{\delta}{\delta \theta_j} p(y_{s_t} | \theta, y^{t-1}, s_{ \leq t}) &= \frac{c_j(y_{s_t} | y^{t-1}, s_{ \leq t})}{Z_t}e^{\theta \cdot c(y_{s_t} | y^{t-1}, s_{ \leq t})} \\ 
    &- \frac{1}{Z_t^2}e^{\theta \cdot c(y_{s_t} | y^{t-1}, s_{ \leq t})}\sum_{i\in{0,1}}{c_j(i | y^{t-1}, s_{ \leq t})e^{\theta \cdot c(i | y^{t-1}, s_{ \leq t})}} \\
    &=\bigg( c_j(y_{s_t} | y^{t-1}, s_{ \leq t}) - E( c_j(Y_{s_t} | y^{t-1}, s_{ \leq t}) | y^{t-1}, s_{ \leq t}) \bigg) p(y_{s_t} | \theta, y^{t-1}, s_{ \leq t})
\end{align*}
}
and
\begin{align*}
    \frac{\delta}{\delta \theta_j} p(y|s,\theta) &= \frac{\delta}{\delta \theta_j} \prod_{t=1}^{n_d} p(y_{s_t} | \theta, y^{t-1}, s_{ \leq t}) \\ 
    &= \sum_{t=1}^{n_d} \bigg(\prod_{k\neq t} p(y_{s_k} | \theta, y^{k-1}, s_{ \leq k}) \bigg) \frac{\delta}{\delta \theta_j} p(y_{s_t} | \theta, y^{t-1}, s_{ \leq t}) \\
    &= p(y|s,\theta) \sum_{t=1}^{n_d} \bigg( c_j(y_{s_t} | y^{t-1}, s_{ \leq t}) - E( c_j(Y_{s_t} \big| y^{t-1}, s_{ \leq t}) | y^{t-1}, s_{ \leq t}) \bigg) \\
    &=p(y|s,\theta) \bigg( g_j(y,s) - \sum_{t=1}^{n_d} E( c_j(Y_{s_t} | y^{t-1}, s_{ \leq t}) \big| y^{t-1}, s_{ \leq t}) \bigg) \\
    &= p(y|s,\theta) \bigg( g_j(y,s) - G_j(y,s) \bigg),
\end{align*}
where $G_j(y,s) = \sum_{t=1}^{n_d} E( c_j(Y_{s_t} | y^{t-1}, s_{ \leq t}) \big| y^{t-1}, s_{ \leq t})$. Note that $E(G_j(Y,S)) = E(g_j(Y,S))$.

\subsection{Relationship to Exponential-family Random Graph Models} \label{sec:ergm}

Exponential-family Random Graph Models are an immensely popular tool for graph analysis, especially in the social sciences \citep{goldenberg2010survey}. Though ERGMs are often presented without a connection to an underlying data generating process, it is popular to interpret the parameters in a way that is consistent with the equilibrium of a tie formation and dissolution process \citep{snijders2006new}.

Similar to the LOLOG model, we begin by positing a sequence ordering $s'$; however, unlike LOLOG, we construct this ordering to be much larger than $n_d$, and each edge variable to have an equal probability of appearing at every time point $p(s'_t = (i,j)) = \frac{1}{n_d}$.

At each time point there is a possibility that the edge variable under consideration will be set to either 0 or 1. Similarly to LOLOG, a logistic relationship is assumed between the change statistics for a set of order independent graph statistics, and the probability distribution for the edge variable under consideration is
\begin{equation}\label{eq:condergm}
p_{\textrm{ERGM}}(y_{s'_t}^t | y^{t-1}, s'_t) = \frac{1}{Z_t}e^{\theta \cdot c(y_{s'_t} | y^{t-1})},
\end{equation}
where the $\theta$ are the natural parameters and 
$$
Z_t = e^{\theta \cdot c(1 | y^{t-1})} + e^{\theta \cdot c(0 | y^{t-1})}
$$
is the normalizing constant.

We see that the functional form for setting edge variables for ERGMs in Equation (\ref{eq:condergm}) is identical to that for LOLOGs in Equation (\ref{eq:condlolog}) except that the ERGM formulation is required to have order independent statistics.

As $t \rightarrow \infty$, the ERGM tie formation and dissolution process reaches a Markov Chain Monte Carlo (MCMC) equilibrium, and the probability of observing a graph is
$$
p_{\textrm{ERGM}}(y^t | \theta) = \frac{1}{Z}e^{\theta \cdot g(y^t)} \ \ \ \ \ y \in \mathscr{Y}
$$
where
$$
Z = \sum_{y'\in \mathscr{Y}} e^{\theta \cdot g(y')}
$$
is a normalizing constant \citep{snijders2006new}. Note that $Z$ involves a sum over all possible graph configurations, which grows exponentially with the size of the network making it difficult or impossible to compute directly for all but the smallest graphs and simplest statistics.

The exception to this is when all $g$ are dyad independent. When this is the case, $Z$ factors into the production of the $Z_t$s, and ERGM reduces to the same logistic regression likelihood as LOLOG.

Because of the logistic formulation, the parameters of an ERGM are interpreted similarly to a logistic regression. Namely, when two actors are ``considering'' forming a tie, a unit increase in $c(1|y^{t-1})$ is associated with an $\theta$ increase in the log odds of an edge being formed.

As $p_{\textrm{ERGM}}$ is the stationary distribution of an MCMC process, other data generating processes may result in the same likelihood. Firstly, $p(s)$ need not be completely at random. Any irreducible proposal may be posited. For example, instead of being completely at random, $s$ could scan deterministic through each edge variable. Secondly, instead of the Gibbs proposal $p(y_{s_t}^t | y^{t-1})$ for edge generation, any transition that follows detail balance would lead to the same stationary distribution. However, choosing a different proposal for the data generating process would complicate the interpretation of the model parameters.

While we have motivated LOLOG via a growth process and ERGM via equilibrium,  both can be thought of as more general than the posited data generating process. They are better thought of as useful languages for distributions on graphs, and it is easy to show that both of them are fully general and able to represent any arbitrary distribution.

\begin{theorem}
For any arbitrary graph distribution $q(y)$:
\begin{enumerate}
    \item There exists a set of graph statistics $g(y)$ and parameters $\theta$ such that
    $$
    p_{\textrm{ERGM}}(y | \theta) = q(y) \ \ \ \ \forall \ \ y \in \mathscr{Y}.
    $$
    \item There exists a set of graph and order statistics $g(y,s)$ and parameters $\theta$ such that for any $p(s)$
    $$
    p(y|\theta) = q(y) \ \ \ \ \forall \ \ y \in \mathscr{Y}.
    $$
\end{enumerate}
\end{theorem}
\textit{Proof:} For (a), we can construct an ERGM for $q$ by setting the parameters equal to the log of the target density at each possible graph configuration, and the statistics equal to an indicator function that the observed graph equals that configuration
\begin{align*}
p_{\textrm{ERGM}}(y | \theta) &= \frac{1}{Z}e^{ \sum_{y^*} \log(q(y^*)) I(y^*=y)}\\
&=\frac{1}{Z} q(y)\\
&= q(y).
\end{align*}

For (b), we focus on the edge formation distribution, and set the parameters equal to the log of a partial conditional distribution of the target, and the change statistics equal to the product of indicators

{\small
\begin{align*}
&p(y_{s_t} | \theta, y^{t-1}, s_{ \leq t}) \\
&=\frac{1}{Z_t} \textrm{exp} \Big( \sum_{i=1}^{n_d}\sum_{s^*_{<i}}\sum_{y^*_{<s^*_i}}\sum_{y^*_{s^*_i} \in \{0,1\}} \log(q(y^*_{s^*_i} | y^*_{<s^*_i})) I(y^*_{s^*_i}=y_{s_t})I(y^*_{<s^*_i} = y_{<s_t})I(s^*_{<i}=s_{<t})I(i=t) \Big) \\
&= \frac{1}{Z_t} q(y_{s_t}| y_{<s_t}) \\
&= q(y_{s_t}| y_{<s_t}). 
\end{align*}
}

Therefore
$$
p(y|\theta) = \sum_s p(s)\prod_{t=1}^{n_d}q(y_{s_t}| y_{<s_t}) = \sum_s q(y)p(s) = q(y).
$$
$\blacksquare$

The first result, that ERGMs can express any distribution is well known, though it is not clear if it has been explicitly described in the literature. Any ERGM distribution can, at least theoretically be expressed as a LOLOG, and vise versa. That said, some network distributions may be parsimoniously represented by an ERGM, some by a LOLOG, and some may not have a parsimonious representation in either.

As the ERGM distribution is an exponential-family, it has several useful characteristics \citep{barndorff1978information}. ERGMs may be parameterized either through the natural parameters, or though the mean value parameters ($\mu = E_\theta(g(Y^t)$). They also have the maximum entropy property, in that they have the largest entropy out of all possible distributions with the same expected sufficient statistics ($E(g(Y^t))$).

Given an observed graph $y^t$, the maximum likelihood estimates for an ERGM ($\hat{\theta}$) satisfy the moment conditions
$$
g(y^t) - E_{\textrm{ERGM}}(g(Y^t) | \hat{\theta}) = 0.
$$
Thus, graphs generated from the MLE distribution have graph statistics $g$ centered around the observed graph statistics. 

Finding the MLE can be challenging due to the general intractibility of $p_{\textrm{ERGM}}$ and hence $E_{\textrm{ERGM}}(g(Y^t) | \theta)$. Typically, the moment conditions are approximated by drawing samples from  $p_{\textrm{ERGM}}$ and then estimating the expectations with the average sufficient statistics in the samples (See \cite{hunter2006inference,snijders2002markov} for details). 

Despite the interpretability and widespread use of ERGM, there are a number of less than desirable features to the model. Simply drawing a sample from an ERGM is itself a non-trivial endeavour as it requires running an MCMC chain to equilibrium. In the best case, where dyads are independent, mixing time is $O(n^2\log(n))$, however, the inclusion of dyad dependency results in asymptotically exponential mixing times \citep{bhamidi2008mixing} as the number of vertices $n$ increases. As a result, considerable, perhaps unattainable, amounts of computational resources are required to estimate parameters from large graphs.

There is also the tendency of many ERGM specifications to display phase transitions (also known as degeneracy) \citep{handcock2003statistical, schweinberger2011instability}. This often manifests as a probability model that puts high mass on the empty and full graphs, with relatively little mass on all other graphs. \cite{chatterjee2013estimating} derived results from large deviation theory suggesting that degenerate states exist asymptotically for a wide array of potential graph statistics and parameter values.


\section{Inference for Order Independent Models via Monte Carlo Method of Moments (MOM)} \label{sec:inf}

In this section we will develop a method of moments estimator for LOLOG parameters given an observed graph in the case of order independent statistics. We adopt the same moment conditions as are typically used in ERGM inference. Namely, we wish to find a $\hat{\theta}$ such that
$$
g(y) - E_{\hat{\theta}}(g(Y)) = 0.
$$
These moment conditions are chosen for three reasons. Firstly, it is desirable to be consistent with ERGM modeling. Secondly, in the case of dyad independence, solving these moment conditions leads to the maximum likelihood estimator. Finally, goodness of fit in networks is often measured by how typical the observed network is on graph statistics compared to those simulated from the fit model \citep{hunter2008goodness}. If these moment conditions are met, the observed graph statistics are centered relative to simulated statistics.

While we can not evaluate $E(g(Y))$ in these moment conditions exactly, due to the intractability of the distribution, we can approximate it by drawing $r$ Monte Carlo samples $(y^{(1)},s^{(1)}),...,(y^{(r)},s^{(r)})$ from $p(y|s,\theta)p(s)$ and estimating $\hat{E}(g(Y)) = \frac{1}{r}\sum_{i=1}^r g(y^{(i)})$.

To perform approximate Newton's method steps, we require the gradient of our moment conditions
\begin{align*}
    D_{jk}(\theta) &= \frac{\delta}{\delta \theta_j} (g_k(y) - E(g_k(Y))) \\
    &= - \frac{\delta}{\delta \theta_j}E(g_k(Y)) \\
    &= - \frac{\delta}{\delta \theta_j} \sum_y \sum_s g_k(y) p(y | \theta, s) p(s) \\
    &= - \sum_y \sum_s g_k(y)\bigg( g_j(y) - G_j(y,s) \bigg) p(y | \theta, s) p(s) \\
    &= -E\bigg(g_k(Y)\bigg( g_j(Y) - G_j(Y,S)\bigg)\bigg) \\
    &= -\textrm{cov}(g_k(Y),g_j(Y)) + \textrm{cov}(g_k(Y), G_j(Y,S)) - E(g_k(Y))E(g_j(Y)) \\
    &\ \ \ +E(g_k(Y))E(G_j(Y,S)) \\
    &= -\textrm{cov}(g_k(Y),g_j(Y)) + \textrm{cov}(g_k(Y), G_j(Y,S)). 
\end{align*}
Like the moment conditions, the covariances in $D$ may be approximated using Monte Carlo sampling. The Newton's method algorithm for solving the moment conditions using a Hotelling's $T^2$ statistic as a termination criteria proceeds as:
\begin{center}
  \uline{A Search Algorithm for MOM Estimation}\\
\end{center}
\begin{enumerate}
    \item Set $k\leftarrow 1$, $\theta^{(k)}$ to some initial values, $r$ to be the number of network samples drawn at each iteration, and $\epsilon>0$ equal to a termination tolerance.
    \item Draw $r$ samples from $p(Y,S|\theta^{(k)})$ and use them in the approximations $\hat{E}_{\theta^{(k)}}(g(Y))$ and $\hat{D}_{jk}(\theta^{(k)})$.
    \item Set $\theta^{(k+1)} \leftarrow \theta^{(k)} - \hat{D}_{jk}(\theta^{(k)})^{-1}(g(y) - \hat{E}_{\theta^{(k)}}(g(Y)))$
    \item If $(g(y) - \hat{E}_{\theta^{(k)}}(g(Y)))^{T}\hat{\textrm{cov}}_{\theta^{(k)}}(g(Y))^{-1}(g(y) - \hat{E}_{\theta^{(k)}}(g(Y))) < \epsilon$, set $\hat{\theta} \leftarrow \theta^{(k)}$ and terminate, otherwise set $k \leftarrow k + 1$ and go to step (b).
\end{enumerate}

After solving the moment conditions to find $\hat{\theta}$, we approximate the covariance of the estimate as
$$
\textrm{cov}(\hat{\theta}) \approx \hat{D}(\hat{\theta})^{-1}\hat{\textrm{cov}}_{\hat{\theta}}(g(Y))\hat{D}^T(\hat{\theta})^{-1}.
$$
Much like parameter covariance estimates for ERGMs, the accuracy of this approximation is somewhat questionable. Developing asymptotic approximations for graphs is challenging in general \citep{chatterjee2013estimating, goldenberg2010survey} due to the fact that only a single graph is typically observed. If multiple graphs were observed, then standard Method of Moments theory shows that a covariance estimate of the above form is accurate as the number of observed graphs becomes large. Additionally, if all terms are dyad independent, then the approximation is accurate for large graphs.

\subsection{Example: Collaboration within a Corporate Law Firm} \label{sec:l}

\cite{lazega2001} collected data on the collaboration connections between partners at a New England law firm. The graph contains 36 vertices, with ties being defined as a pair of individuals both reporting a collaboration relationship with one another. Considerable attention has been paid to modeling this network within the ERGM literature \citep{lazega2001social,van2009framework, hunter2006inference, snijders2006new}. Nodal covariates present in the dataset include seniority (the order of entry into the firm), office (the firm had three offices), practice (litigation or corporate law), and gender.



\begin{table}
\caption{\label{tab:lazega1}ERGM/LOLOG Dyad Independent Model Fit
of the Collaboration Network.} 
\begin{tabular}{rrrr}
  \hline
 & \vline height12pt width0pt\relax $\hat{\theta}$ & Std. Error &  $p$-value \\
  \hline
Edges & -8.31 & 0.95 &  0.00 \\ 
  Seniority(Main Effect) & 0.04 & 0.01 &  0.00 \\ 
  Practice(Main Effect) & 0.90 & 0.16 & 0.00 \\ 
  Gender(Match) & 1.13 & 0.35 & 0.00 \\ 
  Practice(Match) & 0.88 & 0.23 &  0.00 \\ 
  Office(Match) & 1.65 & 0.25 &  0.00 \\ 
   \hline
\end{tabular}

\end{table}

The first model we consider (Table \ref{tab:lazega1}) is a dyad independent model where terms are included for the number of edges (Edges), the sum of the product of the ith and jth value of a covariate over all edges (ij) (Seniority(Main Effect), Practice(Main Effect)), and the number of within group ties (Gender(Match), Practice(Match), Office(Match)). Since this is a dyad independent model, the ERGM and LOLOG likelihoods are identical.

\begin{table}
\caption{\label{tab:lazegaergm}ERGM Fit of the Collaboration Network.}
\centering
\begin{tabular}{rrrr}
  \hline
 & \vline height12pt width0pt\relax $\hat{\theta}$ & Std. Error & $p$-value \\ 
  \hline
Edges & -7.31 & 0.71 &  0.00 \\ 
  GWESP(decay=0.79) & 0.88 & 0.14 &  0.00 \\ 
  Seniority(Main Effect) & 0.02 & 0.01 &  0.00 \\ 
  Practice(Main Effect) & 0.41 & 0.12 &  0.00 \\ 
  Gender(Match) & 0.71 & 0.25 &  0.01 \\ 
  Practice(Match) & 0.76 & 0.19 &  0.00 \\ 
  Office(Match) & 1.14 & 0.19 &  0.00 \\ 
   \hline
\end{tabular}

\end{table}

\cite{hunter2006inference} utilized a geometrically weighted edgewise shared partner (GWESP) term to model the larger than expected number of triangles present in the network. A GWESP term was used because including a triangle term leads to model degeneracy and poor convergence. Table \ref{tab:lazegaergm} summarizes the fit of that model and we see that the GWESP term is positive (0.88) indicating an increased level of transitivity in the network, however, the interpretation of the coefficient beyond that is somewhat difficult given the complex functional form of GWESP.

\begin{table}
\caption{\label{tab:lazegaLOLOG}LOLOG Fit to the Collaboration Network.}
\centering
\begin{tabular}{rrrrr}
  \hline
 & \vline height12pt width0pt\relax $g(y)$ & $\hat{\theta}$ & Std. Error & $p$-value \\ 
  \hline
Edges & 115.00 & -7.65 & 1.07 & 0.00 \\ 
  Triangles & 120.00 & 1.08 & 0.38 & 0.01 \\ 
  Seniority(Main Effect) & 4687.00 & 0.03 & 0.01 & 0.01 \\ 
  Practice(Main Effect) & 359.00 & 0.71 & 0.18 & 0.00 \\ 
  Gender(Match) & 99.00 & 1.15 & 0.44 & 0.01 \\ 
  Practice(Match) & 72.00 & 0.96 & 0.26 & 0.00 \\ 
  Office(Match) & 85.00 & 1.61 & 0.31 & 0.00 \\ 
   \hline
\end{tabular}
 
\end{table}

Table \ref{tab:lazegaLOLOG} shows the result of adding a triangle count term to the dyad independent LOLOG model. The order of vertex inclusion into the network was taken to be the seniority of the lawyer. The triangle term may be interpreted in a natural way. If adding an edge creates an additional closed triangle, then the odds of that edge existing is increased by $e^{1.08} = 294\%$.


\begin{figure}
\centerline{\includegraphics[width=.9\textwidth]{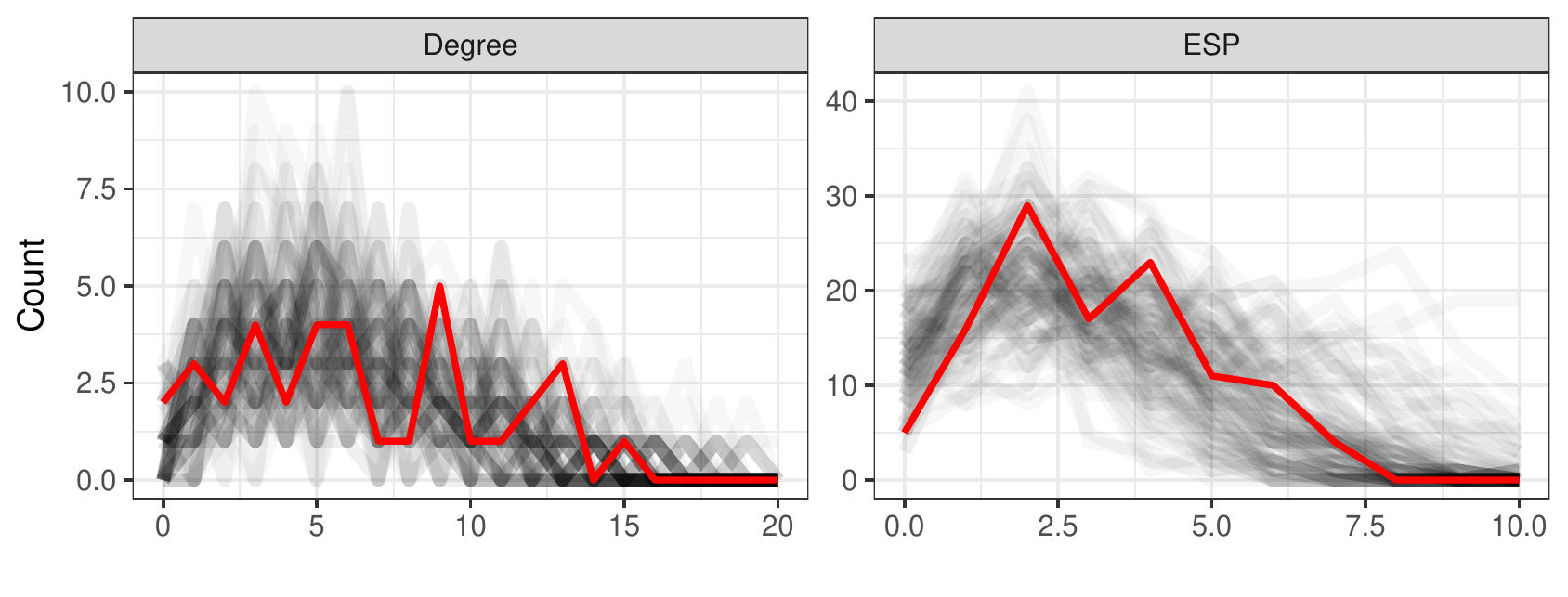}}
\caption{The degree and ESP distributions from 100 simulated networks (black) from the fitted LOLOG compared to the observed collaboration network (\red{red}).\label{fig:lazegagof}}
\end{figure}

One popular method for determining the plausibility of a generative network model is to simulate networks from that model and compare various graph statistics between the simulated networks and the observed network \citep{hunter2008goodness}. If the model is a good fit for the observed network, then the observed graph statistics will be typical of those simulated from the model. Figure \ref{fig:lazegagof} shows the degree distribution and edgewise shared partner (ESP) distribution of simulations from the fitted LOLOG model compared to the observed graph. The observed degree/ESP counts are within the range of the simulated count distributions, and thus the model shows a good fit with regards to these metrics.

\subsection{Example: Friendship Networks Among Adolescent Students} \label{sec:add}

The National Longitudinal Study of Adolescent Health (Add Health) was a stratified survey of US schools containing students from grades 7 to 12. To determine the relationships between friendship connections and health behaviour, students were asked to nominate their peers. Details of this study are outlined in \cite{resnick1997protecting} and \cite{udry1998new}.

Here we analyze one Add Health school, where a friendship tie is defined as two individuals nominating each other. The dataset was restricted to students in grades 9 through 12, yielding a network with 1270 students and 1780 edges. 


The statistics chosen for our LOLOG model are designed to capture some of the major features of this graph. We expect that student grade/gender matters a great deal in tie formation, that students exhibiting the same behaviour (drinking) may be more likely to be tied, and that there will be considerable transitivity (the friend of my friend is more likely to be my friend). Additionally, we will want to model the overall propensity to form ties by way of the degree distribution. 

To do this the following statistics are included in the model. Mixing terms count the number of edges between each grade in the school, and the matching terms count the number of within group edges for sex and alcohol use status. The degree structure was further modeled by a term counting the number of two-stars and one counting the number of vertices with degree $0$. Local clustering was modeled with the inclusion of a triangle count term.

Of particular note is that the LOLOG model utilizes the two-star and triangle terms. These terms are of great interest to network scientists for their interpretability, however, they are rarely used in ERGM due to their propensity for causing model degeneracy in all but the smallest of networks.

The order in which ties are formed is not observed in the network, however, we do observe an approximate partial ordering of students based on their grade. $p(s)$ is specified such that grade 12 students enter the school (and network) first, followed by grade 11, 10, and then 9. Within each grade, the order of inclusion is modeled as random. When a student is added to the network, dyads connecting them to the other individuals already in the network are selected in random order.

\begin{table}
\caption{\label{tab:addhealth}LOLOG Fit to the AddHealth Network} 
\centering
\begin{tabular}{rrrrr}
  \hline
 & \vline height12pt width0pt\relax $g(y)$ & $\hat{\theta}$ & Std. Error & $p$-value \\ 
  \hline
Grade(10-10) & 338.00 & -6.37 & 0.14 & 0.00 \\ 
  Grade(11-10) & 125.00 & -8.08 & 0.17 & 0.00 \\ 
  Grade(12-10) & 34.00 & -9.02 & 0.23 & 0.00 \\ 
  Grade(9-10) & 50.00 & -9.02 & 0.21 & 0.00 \\ 
  Grade(11-11) & 347.00 & -6.68 & 0.15 & 0.00 \\ 
  Grade(12-11) & 110.00 & -8.13 & 0.18 & 0.00 \\ 
  Grade(9-11) & 34.00 & -9.50 & 0.23 & 0.00 \\ 
  Grade(12-12) & 272.00 & -6.54 & 0.18 & 0.00 \\ 
  Grade(9-12) & 8.00 & -10.62 & 0.41 & 0.00 \\ 
  Grade(9-9) & 462.00 & -6.43 & 0.15 & 0.00 \\ 
  Sex(Match) & 1232.00 & 0.78 & 0.05 & 0.00 \\ 
  Drink(Match) & 1149.00 & 0.42 & 0.05 & 0.00 \\ 
  Triangles & 308.00 & 4.57 & 0.13 & 0.00 \\ 
  Degree=0 & 5.00 & -1.21 & 0.16 & 0.00 \\ 
  Two-stars & 4733.00 & -0.05 & 0.02 & 0.04 \\ 
   \hline
\end{tabular}

\end{table}

Table \ref{tab:addhealth} displays the fitted model. The ``Grade'' terms count the number of edges from students in one grade to another grade. We see that the largest parameters are those within grade (9-9, 10-10, 11-11, and 12-12), indicating that students were more likely to be tied to those within their grade. Similarly, the terms where the grade gap is one (9-10, 10-11, 11-12) have larger parameters, but not as large as the within-grade parameters, indicating that ties are likely to form between individuals who are near in age/grade. Similarly, the matching term for gender is positive, indicating ties are more likely between students sharing the same sex.

The parameter for drinking behaviour is positive, so students engaging or not engaging in this are more likely to be tied to each other. It should be noted that drinking behavior is considered exogenous in this model, whereas they likely evolved as the students progressed through high school. An endogenous model would be needed to incorporate this progression. Models of this type have been proposed for ERGM-like models \citep{fellows2012exponential}; however, this type of modeling for LOLOG models is outside the scope of the present paper.

The triangles term is very large, and thus there is a high degree of transitivity, with each potential triangle closure having $e^{4.57} = 57.98$ times higher odds of being closed. Additionally, there are far fewer students with degree 0 than we'd expect by chance, and a negative two-star term indicates that the degree distribution is narrow.

\begin{figure}
\centerline{\includegraphics[width=.9\textwidth]{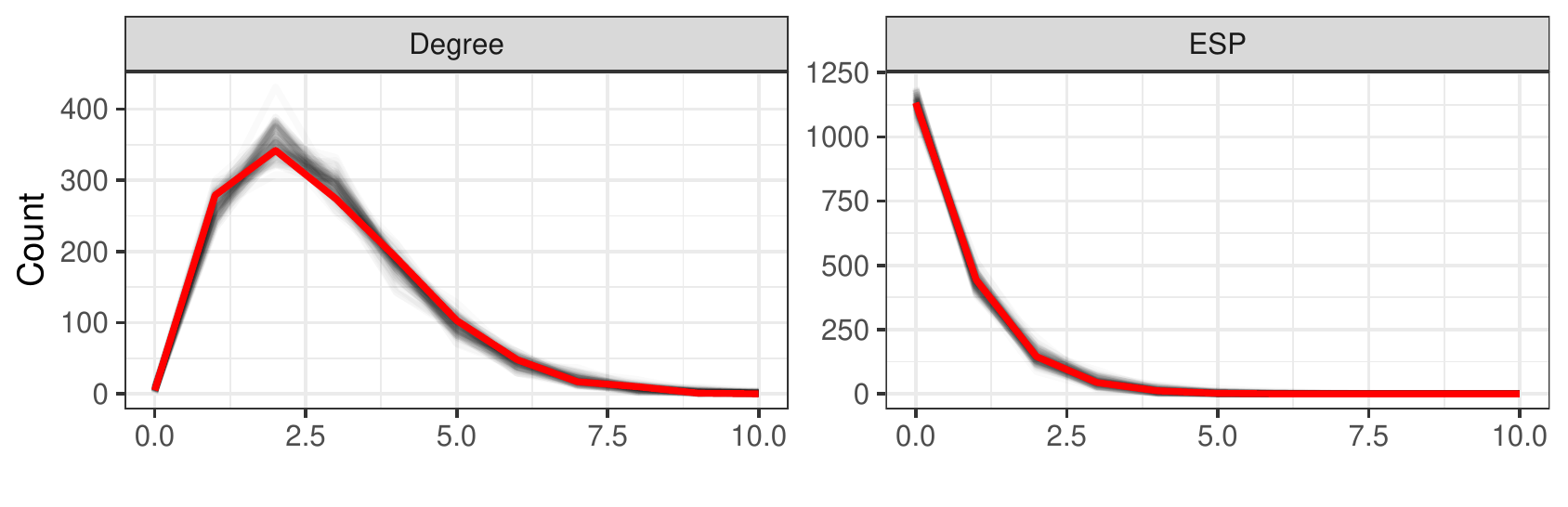}}
\caption{The degree and ESP distributions from 200 simulated networks (black) from the fitted LOLOG compared to the observed Add Health network (\red{red}).\label{fig:add_gof}}
\end{figure}

Figure \ref{fig:add_gof} shows a goodness of fit plot for the degree and ESP distributions in the Add Health network. The networks simulated from the fit model show good agreement with the observed statistics.

Another popular class of generative probability models for networks is the stochastic block model (SBM) \citep{holland1983stochastic}. The simplest formulation posits an unobserved vertex grouping covariate. Conditional upon this grouping variable, ties are formed completely at random. The rate of tie formation within and between each group is controlled by a set of mixing parameters. Conditional upon group membership, an SBM can be thought of as LOLOG/ERGM model with a single mixing term.

Degree corrected SBM models are a popular extension of the basic SBM model that, in addition to the mixing parameters, includes model terms for the degree of each node \citep{karrer2011stochastic}. This allows it to more faithfully represent the degree structure of any observed network, at the cost of model parsimony. This network would seem a natural candidate for a degree corrected SBM due to the grade structure.

\begin{figure}
\centerline{\includegraphics[width=.9\textwidth]{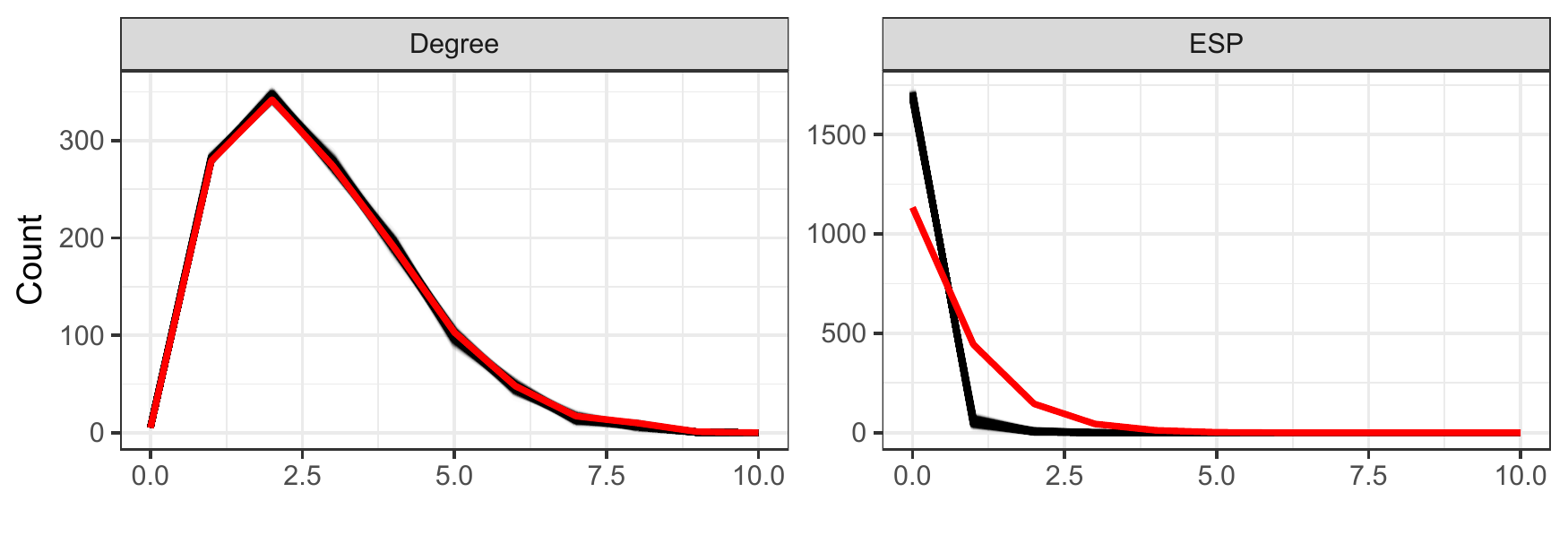}}
\caption{The ESP distributions from 1000 simulated networks (black) from the fitted degree corrected stochastic block model compared to the observed Add Health network (\red{red}).\label{fig:sbm}}
\end{figure}

We fit a degree corrected SBM to the Add Health network using the methods and implementation of \citet{peixoto2014efficient, peixoto_graph}. The algorithm detected seven community blocks. Figure \ref{fig:sbm} shows degree and edgewise shared partner goodness of fit plots for networks generated from the fit SBM. The degree distribution is captured well by the 1270 degree parameters; however, the detected community structure was not enough to capture the local clustering and transitivity present in the graph. Edges in networks simulated from the fitted SBM have far fewer shared partners than exist in the observed graph.

\section{Inference for Order Dependent Models via Monte Carlo GMM-Estimation} \label{sec:inf2}

For models that include order dependent terms, we do not observe the value of the statistic $g(y,s)$ as it depends on the latent order. Thus, direct moment conditions based on the model statistics cannot be used and an alternate means of fitting must be developed.

Let $h(y)$ be a vector of order independent graph statistics and $m(\theta) = h(y) - E(h(Y))$ be a vector of $p \ge |\theta|$ moment conditions. For instance, if we wish to fit the degree distribution, we could include a statistic for each degree equal to the number of vertices with that degree. To model transitivity, we might include a statistic for each possible ESP value equal to the number of edges with that many shared partners.

Generalized Method of Moments \citep{zeger1988models} minimizes the weighted squared deviation of $m(\theta)$
$$
\hat{\theta} = \textrm{argmin}_\theta \ m^{T}(\theta)W m(\theta),
$$
where $W$ is a symmetric positive definite matrix of weights. The first order optimization condition is then
$$
D(\hat{\theta})^{T}Wm(\hat{\theta}) = 0,
$$
where
\begin{align*}
D_{jk}(\theta) = \frac{\delta}{\delta \theta_j}m_k(\theta) &= -\frac{\delta}{\delta \theta_j}E(h_k(y)) \\
&= -E\bigg(h_k(Y)\big( g_j(Y,S) - G_j(Y,S)\big)\bigg) \\
&= -\textrm{cov}(h_k(Y),g_j(Y,S)) + \textrm{cov}(h_k(Y), G_j(Y,S))
\end{align*}                                                                                                                           is the gradient matrix.

A mixture of linearization \citep{newey1985generalized,ahn1997efficient} and Monte Carlo approximations can be used to minimize the criterion function. At each step of the optimization algorithm, we wish to update our previous estimated $\theta^{(k)}$. We begin by substituting in a linear approximation for $m$ centered around the previous values 
\begin{align*}
 m(\theta)' W m(\theta) &\approx \bigg(m(\theta^{(k)}) + D(\theta^{(k)}) (\theta - \theta^{(k)})\bigg)^{T}W \bigg(m(\theta^{(k)}) + D(\theta^{(k)}) (\theta - \theta^{(k)})\bigg). \\
\end{align*}
Differentiating and setting to zero yields the Newton-like update
$$
\theta^{(k+1)} = \theta^{(k)} - \big(D^{T}(\theta^{(k)})WD(\theta^{(k)})\big)^{-1}D^{T}(\theta^{(k)})Wm(\theta^{(k)}).
$$

The following theorem establishes some conditions under which the objective is improved by this update.
\begin{theorem} For positive definite matrix $W$:
\begin{description}
    \item[Descent:] $- \big(D^{T}(\theta^{(k)})WD(\theta^{(k)})\big)^{-1}D^{T}(\theta^{(k)})Wm(\theta^{(k)})$ is a descent direction.
    \item[Monotonicity:] If 
    $$
    \alpha^{-1}D^{T}(\theta^{(k)})WD(\theta^{(k)}) - D^{T}(\theta^{(*)})WD(\theta^{(*)}) - (\nabla^2m(\theta^{(*)}))^{T}Wm(\theta^{(*)})
    $$
    is positive semi-definite for all $\alpha^{*} \in [0,1]$, where $\theta^{(*)} = \theta^{(k)} + \alpha^* (\theta^{(k+1)}-\theta^{(k)})$ and
    $$
    \theta^{(k+1)} = \theta^{(k)} - \alpha \big(D^{T}(\theta^{(k)})WD(\theta^{(k)})\big)^{-1}D^{T}(\theta^{(k)})Wm(\theta^{(k)})
    $$ is the parameter update, then
    $$
    m^{T}(\theta^{(k+1)})W m(\theta^{(k+1)}) \leq m^{T}(\theta^{(k)})W m(\theta^{(k)}),
    $$
    with strict inequality in the case of a definite matrix.
\end{description}
\end{theorem}
\textit{Proof:} 
The first condition is proven by noting that the direction is a descent direction if and only if
$$
\big(D^{T}(\theta^{(k)})Wm(\theta^{(k)})\big)^{T}\big(D^{T}(\theta^{(k)})WD(\theta^{(k)})\big)^{-1}D^{T}(\theta^{(k)})Wm(\theta^{(k)})>0,
$$
which is true if and only if $W$ is positive definite.

The second condition is established following closely Theorem 4.1(a) of \cite{bohning1988monotonicity}. First, let us denote $B=\alpha^{-1}D^{T}(\theta^{(k)})WD(\theta^{(k)})$ and $a=2D^{T}(\theta^{(k)})Wm(\theta^{(k)})$. The update is then $z=-B^{-1}a$ and the second derivative of the objective is
$$
H(\theta) = \nabla^2 m(\theta)^{T}Wm(\theta) =2D^{T}(\theta)WD(\theta) + 2 (\nabla^2m(\theta))^{T}Wm(\theta).
$$
Consider the second-order Taylor expansion around $\theta^{(k)}$
\begin{align*}
    m^{T}(\theta^{(k+1)})W m(\theta^{(k+1)})  - m^{T}(\theta^{(k)})W m(\theta^{(k)}) &= z^{T}a + z^{T}H(\theta^*)z / 2\\
    &\leq z^{T}a + z^{T}B z\\
    &= 0,
\end{align*}
where the inequality is satisfied because $B - H(\theta^*)/2$ is positive semi-definite. \\
$\blacksquare$


The optimal choice of $W$ from an estimation standpoint is $W=\Omega^{-1}$, where $\Omega=\textrm{cov}_\theta(h(Y))$. We utilize the continuously updated GMM estimation approach \citep{hansen1996finite}, setting $W^{(k)} = \hat{\textrm{cov}}_{\theta^{(k)}}(h(Y))^{-1}$ at each step of the algorithm. In order to operationalize this update, we replace $D$ and $m$ by their Monte Carlo counterparts in the following search algorithm:
\begin{center}
  \uline{A Search Algorithm for GMM Estimation}\\
\end{center}
\begin{enumerate}
    \item Set $k\leftarrow 1$, $\theta^{(k)}$ to some initial values, $r$ to be the number of network samples drawn at each iteration, $\alpha^{(1)} \leftarrow 1$, $\beta_1 \leftarrow 0.5$, $\beta_2 \leftarrow 1.2$ and $\epsilon>0$ equal to a termination tolerance.
    \item Draw $r$ samples from $p(Y,S|\theta^{(k)})$ and use them in the approximations $\hat{m}(\theta^{(k)})$ and $\hat{D}_{jk}(\theta^{(k)})$.
    \item If $k>1$ and $\hat{m}^{T}(\theta^{(k)})W^{(k-1)} \hat{m}(\theta^{(k)}) > \hat{m}^{T}(\theta^{(k-1)})W^{(k-1)} \hat{m}(\theta^{(k-1)})$
    \begin{itemize}
        \item Set $\alpha^{(k)} \leftarrow \beta_1 \alpha^{(k-1)}$.
        \item Set $\theta^{(k)} \leftarrow \theta^{(k-1)} - \alpha^{(k)}\big(\hat{D}^{T}(\theta^{(k-1)})W^{(k-1)}\hat{D}(\theta^{(k-1)})\big)^{-1}\hat{D}^{T}(\theta^{(k-1)})W^{(k-1)}\hat{m}(\theta^{(k-1)})$.
        \item Go to Step (b).
    \end{itemize}
    
    \item Set $W^{(k)} \leftarrow \hat{\textrm{cov}}_{\theta^{(k)}}(h(Y))$.
    \item Set $\theta^{(k+1)} \leftarrow \theta^{(k)} - \alpha^{(k)}\big(\hat{D}^{T}(\theta^{(k)})W^{(k)}\hat{D}(\theta^{(k)})\big)^{-1}\hat{D}^{T}(\theta^{(k)})W^{(k)}\hat{m}(\theta^{(k)})$.
    \item If $\hat{D}^{T}(\theta^{(k)})W^{(k)}\hat{D}(\theta^{(k)}) < \epsilon$, set $\hat{\theta} \leftarrow \theta^{(k)}$ and terminate, otherwise set $\alpha^{(k+1)} \leftarrow \textrm{min}(1,\beta_2 \alpha^{(k)})$, $k \leftarrow k + 1$ and the go to Step (b).
\end{enumerate}
Since $W^{(k)}$ is a sample covariance matrix it is positive semi-definite, nearly meeting the criteria for ensuring a descent direction. If the matrix is indefinite, then the update step will fail. However, this will only occur in practice if some of the $h$ are linearly dependent and may be avoided by adding a positive diagonal matrix to the covariance matrix.

GMM asymptotic theory provides a nominal parameter covariance matrix of
$$
\textrm{cov}(\hat{\theta}) \approx (\hat{D}^{T}W\hat{D})^{-1}\hat{D}^{T}W\hat{\Omega} W^{T}\hat{D}(\hat{D}^{T}W^{T}\hat{D})^{-1},
$$
where $\hat{\Omega} = \hat{\textrm{cov}}(h(Y))$. As with ERGM parameter covariances and LOLOG order independent estimates, the asymptotic frame for this is somewhat questionable since we only observe a single network.

\section{Algorithmic Initialization and Fast Approximate Inference} \label{sec:var}

It is desirable to have a fast method of obtaining LOLOG parameter estimates. This allows for approximate inference to be preformed on larger networks where the MOM and GMM approaches may be insufficiently fast. Perhaps more importantly, these parameter values can be used as starting values for MOM or GMM estimation, leading to faster convergence of those algorithms.

We begin with the variational approximation for the log likelihood given an arbitrary density over the dyad inclusion order $q(s)$
\begin{align*}
    \log p(y | \theta) &= \sum_s \log p(y | \theta) q(s) \\
    &= \sum_s \log\bigg( \frac{p(y,s | \theta) / q(s)}{p(s | y, \theta) / q(s)}\bigg)q(s) \\
    &\propto E_q\bigg(\log\big(\frac{p(y,S|\theta)}{q(S)}\big)\bigg) + E_q\bigg(\log\big(\frac{q(S)}{p(S | y, \theta)}\big)\bigg) \\
    &\ge E_q\bigg(\log\big(\frac{p(y,S|\theta)}{q(S)}\big)\bigg).
\end{align*}
Variational inference replaces the maximization of the log likelihood, by the bounded expression in the last equation. 

The variational approximation is exact if $q(s) = p(s | y, \theta)$, however, this distribution is generally intractable and difficult to sample from. A more convenient distribution is $q(s) = p(s)$, which we can sample from easily. Additionally, if the model is dyad independent, then $p(s) = p(s | y,\theta)$ and the variational solution is identical to the maximum likelihood solution. We wish to maximize the function
\begin{equation}
    Q(y,\theta) = E_{p(S)}\big( \log p(y | S, \theta)\big),
\end{equation}
and will approximate it using Monte Carlo sampling.

Let $s^{(1)},...,s^{(r)}$ be independent samples from $p(s)$, then
\begin{align} \label{eq:var}
    \hat{Q}(y,\theta) &= \frac{1}{r}\sum_{k=1}^r \log p(y | s^{(k)}, \theta) \nonumber \\
    &= \frac{1}{r}\sum_{k=1}^r \sum_{t=1}^{n_d}  \bigg(\theta \cdot c(y_{s_t} | y^{t-1}, (s_{ \leq t})^{(k)}) - \log Z_t^{(k)}\bigg).
\end{align}
Conveniently, we can recognize that Equation (\ref{eq:var}) is a logistic regression criterion function with the change statistics as predictors and the dyads as outcomes. So the variational estimates may be obtained by maximizing $\hat{Q}$
$$
\hat{\theta}_V = \textrm{argmax}_\theta \ \ \hat{Q}(y,\theta),
$$
which can be done using standard logistic regression fitting software.


\section{Using LOLOG to Model Scale-Free Networks via Preferential Attachment} \label{sec:scale}

It is often claimed that empirical networks display degree distributions that have power law behavior; in that the degree distribution is approximately
$$
p(d) \propto d^{-\gamma}
$$
for $d> M$, where $M$ is some minimum degree \citep[See][and the references therein]{barabasi2016network}. This claim is controversial \citep{joneshandcock2003a,joneshandcock2003b,handcockjones2004, clauset2009power, stumpf2012critical}.

One popular algorithm for generating networks with a power law distribution is the Barabasi-Albert model \citep{barabasi1999emergence}. Under this model we begin with an initial connected network of nodes. Nodes are then added to the network one at a time. Each connects with $m$ existing nodes with probability proportional to their degree (possibly plus a constant $k$).
$$
p_i = \frac{k+d_i}{\sum_j (k+d_j)}.
$$

The preferential attachment mechanism of the Barabasi-Albert model results in rich nodes (those with high degree) getting even richer at the expense of low degree nodes, and has a degree distribution that approximately follows a power law with $\gamma=3$ as the network grows larger.

Inspired by this model, we can design a LOLOG model that exhibits similar behavior to the Barabasi-Albert model. Consider a model with two terms, with the first term being an edges term and the second term representing preferential attachment. If the node ``entering'' the network is known as the acting node and $a_t$ is the alter of that node under consideration at time $t$, then we define the preferential attachment statistic when an edge is present as
$$
c_2(y_{s_t}=1 |y^{t-1},s_{\leq t}) = \log \frac{k + d^{t-1}_{a_t}}{ \sum_{j}(k + d^{t-1}_j) }.
$$
The connection between the Barabasi-Albert model and the LOLOG model is apparent when we examine the probability of forming an edge
\begin{align*}
p(y_{s_t} = 1 | \theta, y^{t-1}, s_{ \leq t}) &= \frac{e^{\theta \cdot c(y_{s_t} = 1 | y^{t-1}, s_{ \leq t})}}{1 + e^{\theta \cdot c(y_{s_t}=1 | y^{t-1}, s_{ \leq t})}} \\
&\approx e^{\theta \cdot c(y_{s_t} = 1 | y^{t-1}, s_{ \leq t})} \\
&=e^{\theta_1} \Bigg( \frac{k + d^{t-1}_{a_t}}{ \sum_{j}(k + d^{t-1}_j) } \Bigg)^{\theta_2}.
\end{align*}
The approximation is valid when the probability of forming the edge is low, which is generally true in sparse graph models. When $\theta_2=1$ and $\theta_1=0$ we see that the probability of forming an edge is nearly identical to that of the Barabasi-Albert model. The expected number of edges added by each node entering the network is equal to the sum of these probabilities from the time when the node entered to when the next node enters. Let $u_i$ be the time the ith node enters, then
$$
E(d_i^{u_{i+1}-1}) \approx \sum_{u_i \leq t < u_{i+1}} e^{\theta_1} \Bigg( \frac{k + d^{t-1}_{a_t}}{ \sum_{j }(k + d^{t-1}_j) } \Bigg)^{\theta_2} \approx e^{\theta_1} \frac{\sum_{j}(k + d^{u_i-1}_{a_t})^{\theta_2}}{ \big(\sum_{j}(k + d^{u_i-1}_j)\big)^{\theta_2} }.
$$
For $\theta_2=1$ the expected number of edges added by each node simplifies to $e^{\theta_1}$. If $\theta_2>1$, then by Jensen's inequality, we would expect the number of edges created to decrease over time, while when $\theta_2 < 1$ we would expect them to increase.

\begin{figure}
\centerline{\includegraphics[width=.9\textwidth]{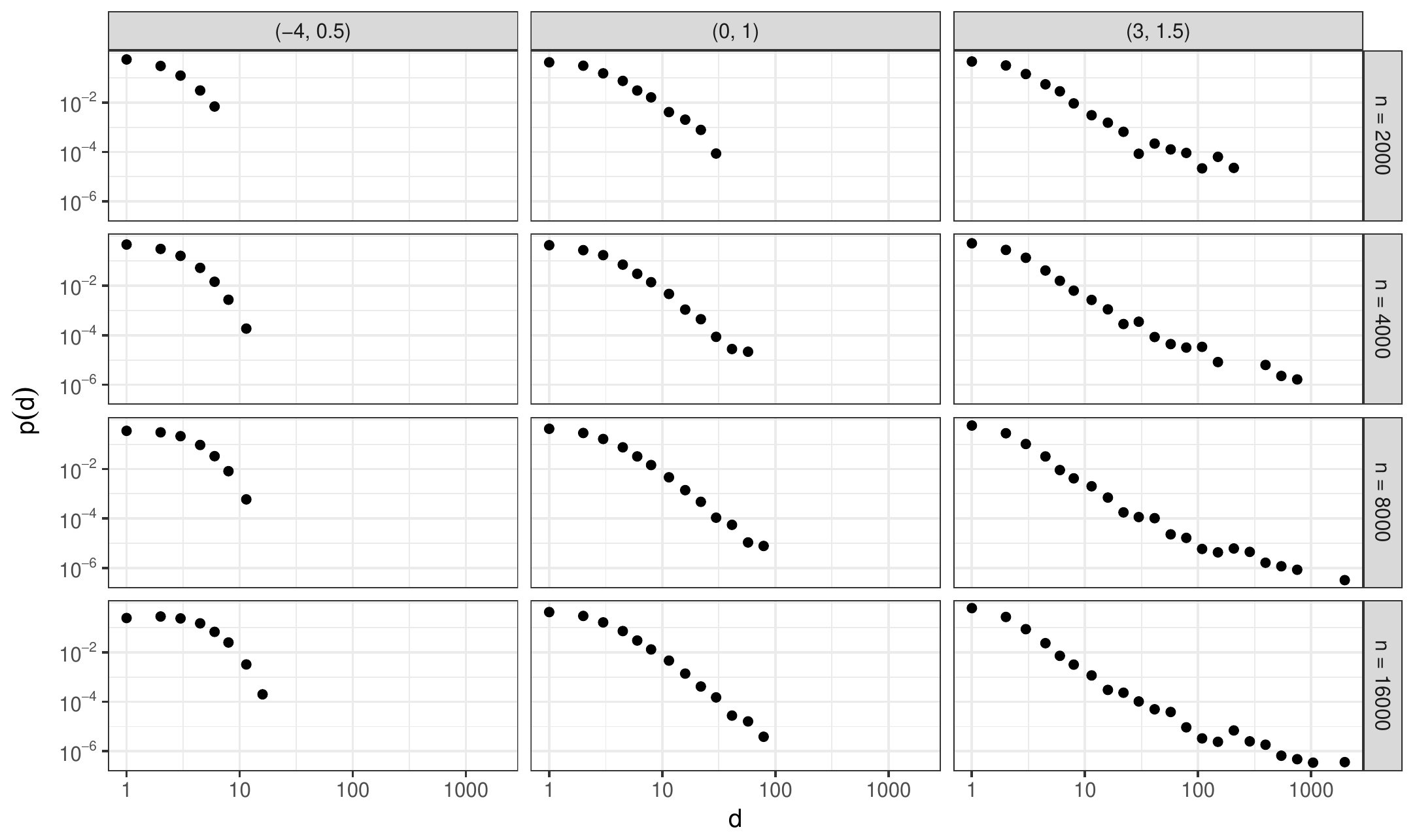}}
\caption{The log-log degree distribution of 12 simulated networks from the preferential attachment model with $k=1$ and with different network sizes and parameter values. The left panel has sublinear ($\theta_2=.5$), the middle panel has linear ($\theta_2=1$) and the right panel has superlinear ($\theta_2=1.5$) degree growth. Degrees are binned in the log scale.\label{fig:scale}}
\end{figure}

Figure \ref{fig:scale} shows the degree distribution of the LOLOG preferential attachment model for different network sizes and parameter values and Table \ref{tab:scale} displays the mean degrees for the network. For $\theta=(0,1)$ we see good agreement with our expectations. The mean degree is constant over time and matches the values expected from the approximations (mean degree $=2e^0=2$). The degree distribution is linear on the log-log scale, and is stable over different network sizes. For $\theta_2=1.5$ we see super-linear behavior in the log-log scale, with mean degree decreasing over time. Finally, when $\theta_2=0.5$ the degree distribution is sublinear and the mean degree increases over time. Interestingly, the relationship between the increase in mean degree and network size is very close to linear. This sort of densification of the graph has been seen in analyses of real networks \citep{leskovec2005graphs}.

\begin{table}
\caption{\label{tab:scale}Mean degree in 12 simulated graphs from the preferential attachment model with different network sizes and parameter values.}
\centering
\begin{tabular}{r|rrr}
  \hline
 \vline height12pt width0pt\relax $n \backslash \theta$ & $(-4,0.5)$ & $(0,1)$  & $(3,1.5)$ \\ 
  \hline
2000 & 1.06 & 2.03 & 2.59 \\ 
  4000 & 1.46 & 1.96 & 2.20 \\ 
  8000 & 2.07 & 2.02 & 1.91 \\ 
  16000 & 2.94 & 1.98 & 1.56 \\ 
   \hline
\end{tabular}

\end{table}

\section{Application: Modeling the Social Relations of Hamsters (and their Guardians)} \label{sec:ham}

Hamsterster.com was once the premier location for hamsters and/or their caretakers to engage in online social networking. Caretakers could create profiles for their pets, and add friendship links to other hamsters. Sadly, the site was shuttered in 2014 after nearly 10 years of operation, leaving this population critically underserved by the social networking community \citep{dunker2015social}.

Here we analyze the friendship relations of the Hamsterster network as it existed in 2012 \citep{kkonest, dunker2015social, petsterfriendshipshamster}. There are a total of 1856 hamster profiles in the network with 12,534 undirected edges. Hamsters with no friendship connections, or whose profiles were empty were not included in the dataset, thus the minimum degree in the network is 1. 

Profiles contain various descriptions of the hamsters and their preferences. The date the hamster joined the network is included, and is therefore used in our LOLOG modeling as an ordering variable. Additionally, the location of the hamster is specified by hometown, state and country. For analysis we split this into two hierarchies. The locale is defined as the state or province that the hamster is located in, or the country if it is not in the United States or Canada. Town is defined as the hometown of the hamster. Fur coloration is also included with Grey and Black being most prevalent. A number of different hamster ``species'' are present in the dataset, with Syrian and Dwarf being the most common. Users also were allowed free-form text spaces to describe the hamsters' favorite toys and food. For analysis these were simplified by replacing the free-form text with common hamster ``likes,'' with less common likes overriding more common likes when a hamster displayed a predilection for multiple things. For the favorite toy variable, hamsters most commonly liked wood, wheel, chew, cardboard, ball, tubes, cage and house. For food, they often liked drops, seed, yogurt, lettuce, peanut, fruit, strawberry, banana, carrot, cheese, apple and corn. To model the density of the network over time, a term was added equal to the log of the order of the active vertex. This is an order independent term because we have observed the order of individuals entering the network through their join date.

In addition to nodal covariates two order dependent terms were included in the model. The Hamsterster network displays a long tailed degree distribution, and thus the Preferential Attachment term of the previous section is included. Instead of including a raw triangle term, an order dependent transitivity term is included, where the change statistic for an added edge is the log number of shared neighbors (at the current time) divided by the maximum potential number of shared neighbors, which is the minimum of the degrees of the nodes connected by the edge. Since we wish the log to be defined when there are no shared edges, one is added to the ratio inside the log. This metric is a more reasonable measure of transitivity for networks with high degree variability, as we would expect much larger amounts of triangles to be formed between high degree nodes, than between low degree nodes, even in the absence of clustering. For GMM estimation all of the order independent statistics in the model are included as $h$ statistics. Additionally, the number of two-stars and Soffer-Vasquez transitivity \citep{soffer2005network} were included as $h$ statistics to model the degree spread and clustering of the network.

\begin{table}
\caption{LOLOG fit for Hamsterster network\label{tab:ham}}
\centering
\begin{tabular}{rrrrr}
  \hline
 & \vline height12pt width0pt\relax $h(y)$ & $\hat{\theta}$ & Std. Error & $p$-value \\ 
  \hline
  &Order Independent\\
  \hline
Edges & 12534.00 & -0.16 & 0.29 & 0.59 \\ 
  Food(Match) & 688.00 & 0.12 & 0.09 & 0.21 \\ 
  Toy(Match) & 1944.00 & 0.29 & 0.07 & 0.00 \\ 
  Species(Match) & 4001.00 & 0.53 & 0.05 & 0.00 \\ 
  Color(Match) & 882.00 & 0.29 & 0.06 & 0.00 \\ 
  Town(Match) & 351.00 & 0.80 & 0.16 & 0.00 \\ 
  Locale(Match) & 1184.00 & 0.87 & 0.06 & 0.00 \\ 
  Degree=1 & 284.00 & 2.28 & 0.07 & 0.00 \\ 
  log(Order) & 82047.44 & 0.17 & 0.04 & 0.00 \\ 
  \hline
  &Order Dependent\\
  \hline
  Preferential Attachment & NA & 0.88 & 0.06 & 0.00 \\ 
  Shared Neighbors & NA & 4.38 & 0.40 & 0.00 \\ 
  \hline
  &Moment Conditions\\
  \hline
  Two-stars & 555847& \\
  SV-Transitivity & 0.194& \\

   \hline
\end{tabular}
\end{table}

Table \ref{tab:ham} shows the fit model parameters. We see that hamsters (or their owners) display a propensity to have ties with those hamsters that share similar toys, or are of the same species or color. Geographic location also plays a large role, with highly significant matching terms for town and local. Food choices do not have a significant impact on tie formation. The preferential attachment term is large, but more than two standard deviations below one, possibly indicating a sub-linear (in the log-log scale) degree distribution. The shared neighbor term is highly significant, so transitivity is high. The log(Order) term is positive, so the degrees over time are higher than would be expected given the other terms. Interestingly, the edges term is not significant, so the other terms are sufficient to explain the overall density of the network.

\begin{figure}
\centerline{\includegraphics[width=.9\textwidth]{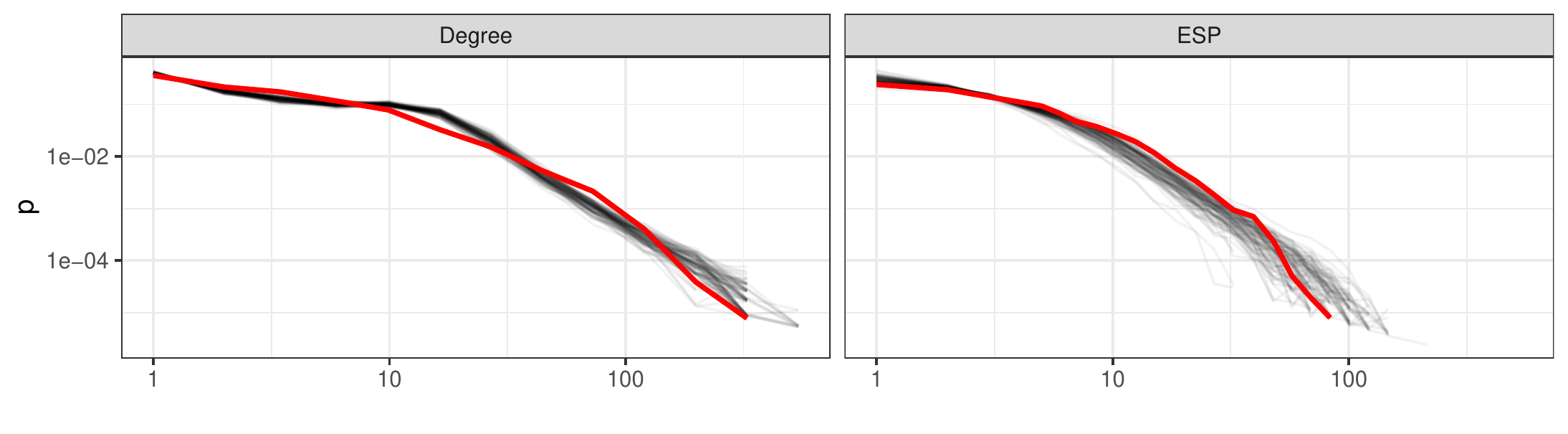}}
\caption{The log-log degree/ESP distribution of 100 simulated networks from the fitted model and the observed Hamsterster network (\red{red}). \label{fig:hamdeg}}
\end{figure}

\begin{figure}
\centerline{\includegraphics[width=.7\textwidth]{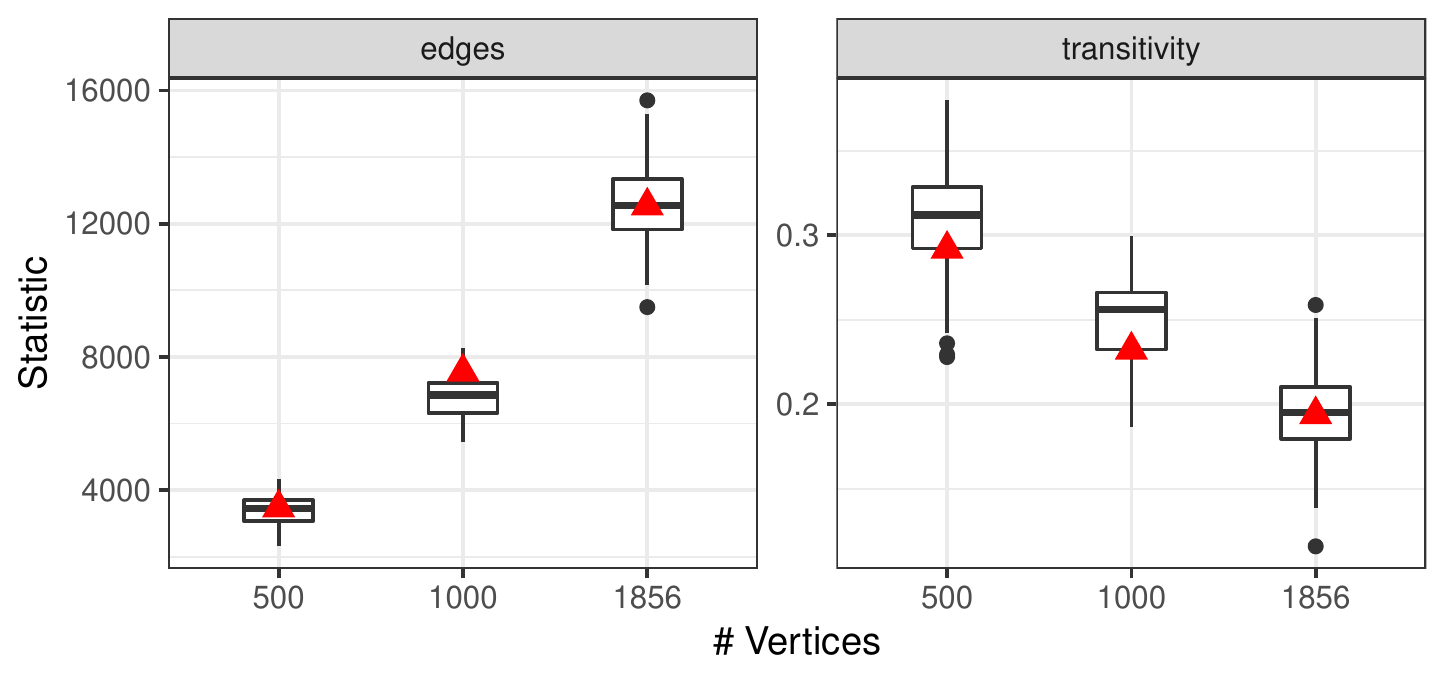}}
\caption{The distribution of \# of edges and network transitivity as the network grows for 100 simulated networks from the fitted model. The growth of the observed Hamsterster network is marked in red. \label{fig:hamgrow}}
\end{figure}

Figure \ref{fig:hamdeg} shows a goodness of fit for the degree and ESP distributions. Unlike the previous goodness of fit plots, these are binned and displayed on the log-log scale due to the long tails of the distributions. The ESP distribution of the observed network is typical of those seen in simulated networks from the model. The degree distribution fits relatively well, with perhaps a bit too much linearity in the tail of the distribution.

Figure \ref{fig:hamgrow} shows a goodness of fit plot for the growth of the network relative to the Soffer-Vasquez transitivity and number of edges. The red triangles show the statistics for networks limited to the first 500, 1000 and 1856 individuals to enter the network based on their sign-up date. The boxplots show simulated networks of those sizes from the model. The observed network looks fairly typical with the number of edges increasing and transitivity decreasing in line with the simulated networks.

\section{Discussion}

The structure and formation of empirical networks are generally quite complex and thus they require flexible statistical modeling tools to describe and analyze faithfully. The LOLOG framework represents a new perspective on network modeling, one that is motivated by the idea of network growth, and yet is general enough to be capable of describing any probability distribution on graphs.

LOLOG can be seen as complimentary to the popular ERGM framework. Both are fully general, so distributions expressible as ERGMs can also be expressed as LOLOG. That said, some distributions may be parsimoniously described by LOLOG with just a few terms, whereas an ERGM might require as many terms as there are graph configurations to reproduce the same distribution. The reverse is also true, with simple ERGM models being impractical to reproduce in LOLOG. Finding cases where there is a computable and understood mapping between LOLOG and ERGM models is an open problem. However, in the case where all terms are dyad independent, there is an exact correspondence between LOLOG and ERGM.

Due to their generality both ERGM and LOLOG are in theory compatible with any underlying graph data generation process, though their motivating data generating processes are quite different. LOLOGs are motivated by network growth, where edge variables are evaluated sequentially and remain static after consideration. ERGMs on the other hand are motivated through the equilibrium of a tie formation/dissolution process on a graph of static size. The plausibility of either of these processes as an approximation of the true data generating process is likely to be domain dependent.

Each of the networks analyzed in this paper contained some information that could plausibly inform the growth process. For the collaboration network within a corporate law firm, we utilized the fact that senior members existed at the firm prior to junior members. Similarly, for the Add Health school, friendships between those in higher grades are considered to predate those in lower grades. For the Hamsterster network, order of inclusion into the network was directly collected. Despite this information, the LOLOG growth process is necessarily an approximation. Some friendships in the Add Health high school almost certainly dissolved, with new friendships formed in their place. Lawyers in the Lazega firm almost certainly left for other firms, leaving them unobserved.

From a practical standpoint LOLOG models like the ones in our examples, show considerable model stability, avoiding the computational difficulties of MCMC sampling, and the modeling challenges posed by phase transitions. This allowed the use of terms like two-stars and triangles without causing model degeneracy, improving our ability to utilize interpretable graph statistics.

Numerous examples exist in the literature of proposed network growth procedures; however, they have typically been limited to modeling a single feature or set of network datasets \citep[e.g.,][]{barabasi1999emergence, leskovec2005graphs, krapivsky2000connectivity, holme2002growing}). Often, growth procedures are presented without a fully specified stochastic model, and with no mechanism for performing statistical inference. What has been lacking is a general framework from which to view different growth procedures, sample from them, and perform inference. In Section \ref{sec:scale} we saw how the classic Barabasi-Albert model was quite naturally expressible in LOLOG terms. It is an open and interesting question as to what other models in the literature are easily expressible in LOLOG terms, and how these existing models might motivate a suite of graph statistics that can be combined and selected as needed by research practitioners.

\bibliographystyle{apalike}
\bibliography{lolog}

\end{document}